\documentclass[conference]{IEEEtran}

\PassOptionsToPackage{table,dvipsnames}{xcolor}

\usepackage{newtxtext,newtxmath}

\usepackage{amsmath,bm}

\usepackage{graphicx}
\usepackage{algorithm}
\usepackage{algorithmic}
\usepackage{textcomp}
\usepackage{comment}
\usepackage{pifont}

\usepackage{xcolor}

\usepackage{hyperref}

\def\BibTeX{{\rm B\kern-.05em{\sc i\kern-.025em b}\kern-.08em
    T\kern-.1667em\lower.7ex\hbox{E}\kern-.125em X}}

\newcommand{\name}{MAHL}

\definecolor{lightgray}{gray}{0.9}
\definecolor{lightgreen}{rgb}{0.85,1.0,0.85}

\begin{document}

\title{MAHL: Multi-Agent LLM-Guided Hierarchical Chiplet Design with Adaptive Debugging
}
\author{
\IEEEauthorblockN{
Jinwei Tang\IEEEauthorrefmark{2}, 
Jiayin Qin\IEEEauthorrefmark{2}, 
Nuo Xu, 
Pragnya Sudershan Nalla, 
Yu Cao, 
Yang Katie Zhao, 
Caiwen Ding
}
\IEEEauthorblockA{
\textit{University of Minnesota - Twin Cities}}
Email: tang0940@umn.edu, qin00162@umn.edu, xu001536@umn.edu, nalla052@umn.edu, \\yucao@umn.edu, yangzhao@umn.edu, dingc@umn.edu
\thanks{\IEEEauthorrefmark{2} Co-first authors.} 
}

\maketitle

\begin{abstract}
As program workloads (e.g., AI) increase in size and algorithmic complexity, the primary challenge lies in their high dimensionality, encompassing computing cores, array sizes, and memory hierarchies.
To overcome these obstacles, innovative approaches are required. Agile chip design has already benefited from machine learning integration at various stages, including logic synthesis, placement, and routing. 
With Large Language Models (LLMs) recently demonstrating impressive proficiency in Hardware Description Language (HDL) generation, it is promising to extend their abilities to 2.5D integration, an advanced technique that saves area overhead and development costs. However, LLM-driven chiplet design faces challenges such as flatten design, high validation cost and imprecise parameter optimization, which limit its chiplet design capability.
To address this, we propose MAHL, a hierarchical LLM-based chiplet design generation framework that features six agents which collaboratively enable AI algorithm-hardware mapping, including hierarchical description generation, retrieval-augmented code generation, diverseflow-based validation, and multi-granularity design space exploration. These components together enhance the efficient generation of chiplet design with optimized Power, Performance and Area (PPA). Experiments show that MAHL not only significantly improves the generation accuracy of simple RTL design, but also increases the generation accuracy of real-world chiplet design, evaluated by Pass@5, from 0 to 0.72 compared to conventional LLMs under the best-case scenario. Compared to state-of-the-art CLARIE (expert-based), MAHL achieves comparable or even superior PPA results under certain optimization objectives.
\end{abstract}


\section{Introduction}
\label{sec:intro}
 The slowing of Moore's law and increasing demand for AI-specific Integrated Circuits (IC), have driven interest in developing new hardware solutions and computing paradigms that deliver high performance and energy efficiency~\cite{Hennessy2019,Theis2017,Bohr2019}.
2.5D integration, an advanced packaging technique, has gained popularity due to its scalability and modularity to meet the growing computational demands of AI workloads~\cite{Simba,UCIE,EMIB,interposer}. Typically, 2.5D IC interconnects multiple chiplets,
either homogeneously or heterogeneously, 
within a single package via an interposer, thereby alleviating limitations for monolithic accelerators in area overhead and development costs. 

To realize the 2.5D integration that accommodates different AI applications, the design of reusable and configurable chiplet IPs remains an indispensable and time-consuming venture. The inefficiency of the existing chiplet IP design methodology can be mainly attributed to two reasons. First, manually designing, modifying, and verifying hierarchical and reusable hardware IPs demands substantial design time and expertise. Furthermore, the large-scale AI applications significantly expand design space, which poses major challenge to both design efficiency and quality in the chosen design. Therefore, to address these limitations, we need an efficient approach to optimize the generation flow of chiplets, not only enabling the rapid, cost-effective development of high-performance hardware designs, but also leveraging efficient DSE to identify the optimal configuration for specific AI applications.

Large Language Models (LLMs) have emerged as a promising solution in various hardware-related tasks. With a Multi-Agent design, LLM accuracy and stability are further enhanced. Prior works have demonstrated their potential in automatically generating HDL code from natural language descriptions or high-level specifications~\cite{thakur2023benchmarking,MG-Verilog,Chipgpt,Chateda}. Some studies~\cite{GPT4AIGCHip,vungarala2024sa} have further explored the design space with the assistance of LLMs. Inspired by their potential, we aim to integrate LLMs into the traditional chiplet generation flow. However, through analyzing LLM-generated hardware designs, we observe three critical issues for integration: (1) \textit{Flatten designs}: LLMs tend to generate all code within one single block, which fails to meet the modular requirements of chiplet design; (2) \textit{High validation cost}: LLM-generated HDL code often lacks sufficient accuracy, while manual effort to develop and verify testbenches remains high;  (3) \textit{Imprecise parameter optimization}: LLMs still struggle with the precise optimization of configuration parameters~\cite{HLSPilot}, making it challenging to directly determine the most efficient configurations. 

Based on these observations, we propose~{\name}, an LLM-guided chiplet IP generation framework that decomposes chiplet generation tasks into LLM-manageable submodules and further reaps the hierarchical structures for DSE optimization. The contributions of our proposed~{\name} can be summarized as follows:


\begin{itemize}

\item 
We propose a Hierarchical Parser and a Hierarchical Module Description Generator for AI-hardware co-design, incorporating two mechanisms. The AI-hardware mapping mechanism enables automated RTL generation by translating high-level AI models into hardware modules, while the hierarchical prompt splitting mechanism enhances module reuse, aligning with the modular and reusable nature of chiplet-based architectures.

\item 
MAHL incorporates a Retrieval-Augmented Code Generator that leverages dynamic module selection to reuse correct and high-quality modules, thereby improving generation efficiency and promoting module reuse. In addition, our framework integrates a Diverseflow Validator that injects controlled noise into the debugging process to enhance output diversity, effectively preventing the validation and correction loop from getting stuck on a single error case.

\item 
For the DSE in MAHL, we design a Multi-Granularity Design Space Explorer that integrates the strengths of both LLMs and analytical techniques. It employs LLMs to perform a coarse-grained, breadth-first exploration over the large design space, followed by analytical DSE for fine-grained, depth-oriented refinement to identify locally optimal configurations. This hybrid approach, together with a reinforcement-driven feedback loop guided by bottleneck analysis, leverages the domain knowledge of LLMs to efficiently navigate and optimize within an expansive design space.

\end{itemize}

The MAHL framework, leveraging LLM-generated hierarchical prompts, improves average Pass@1 success rates by $44.67\%$ over conventional methods in simple RTL designs, achieving up to a 0.72 Pass@5 increase for complex AI designs. Compared to human-designed chiplets, MAHL-generated designs show an average latency reduction of 16.08\% in high-performance mode and an 83.96\% area reduction in compact-area mode, highlighting MAHL's significant potential in chiplet generation.

\section{Background}
\label{sec:rw}

\subsection{AI Workload}
AI algorithms have experienced an explosive growth over the past decades.  The classical Convolutional Neural Networks (CNNs) have played a pivotal role in advancing computer vision, leading to significant model innovations~\cite{VGG16, ResNet}. As research progresses, its focus further expands to Natural Language Processing (NLP), driving the evolution of LLMs with increased diversity and complexity. For instance, GPT~\cite{GPT} leverages data-driven pretraining and multitask learning to address a wide range of NLP tasks, while LLaMA~\cite{Llama}, an open-source auto-regressive model, optimizes for performance with limited resources. BERT~\cite{bert}, on the other hand, specializes in bidirectional context understanding. Therefore, in this paper, we select three representative LLM models, BERT~\cite{bert}, GPT~\cite{GPT}, and LLaMA~\cite{Llama}, as our target algorithms.

\subsection{Chiplet Design Flow}


Prior research~\cite{chopin,Heterogenous} utilize custom frameworks to propose the chiplet-based design, leveraging its advantages in scalability and modular reuse. For example, the Chopin method utilizes reusable algorithmic chiplets, enabling flexible combination of matrix multiplication and other computational modules based on task requirements, thereby improving hardware resource utilization~\cite{chopin}. 

Inspired by prior work, we outline a four-phase chiplet design flow: (1) Design Specification, (2) Behavior Modeling, (3) Design Space Exploration and RTL Implementation, and (4) Physical Layout. \textbf{Phase I} defines system-level specs tailored to target algorithms. \textbf{Phase II} builds behavioral models for early performance/resource estimation. \textbf{Phase III} conducts DSE to identify optimal architectures, followed by final RTL synthesis. Though both Phases II and III involve RTL, the former aids modeling, while the latter finalizes the design. \textbf{Phase IV} performs backend tasks like place-and-route and DRC/LVS checks.

To streamline the labor-intensive module design and DSE phases, we propose integrating LLMs to automate modular, configurable, and reusable chiplet generation.

\subsection{LLM-aided Hardware Generation}
Traditional hardware design utilizes Electronic Design Automation (EDA) tools to streamline the flow, thereby reducing the need for manual intervention~\cite{ Gemmini,  DSAGEN, RF-CGRA}. However, with the evolution of LLMs, researchers have explored more automatic hardware generation and debugging flow~\cite{AI-Native-EDA}, leveraging techniques like prompt engineering and fine-tuning~\cite{VerilogEval, RTLCoder, RTLLM}.

Recent works such as Chip-Chat~\cite{blocklove2023chip} and RTLCoder~\cite{liu2024rtlcoder} demonstrate the potential of LLMs in RTL generation, while VGV~\cite{vgv2024} enhances Verilog synthesis. However, scaling to complex, hierarchical designs remains challenging. 

To address this issue, \cite{GPT4AIGCHip} proposes the GPT4AIGChip framework, which utilizes a demo-augmented pipeline to automate template-based AI accelerator design with LLMs. 
ROME~\cite{ROME} introduces an automatic hierarchical generation pipeline without human feedback, but it lacks support for AI-chiplet design, prioritizing generation accuracy over key aspects like module reuse and parameterized PPA optimization.

 \begin{figure*}[t]
     \centering
\includegraphics[width=1.0\linewidth]{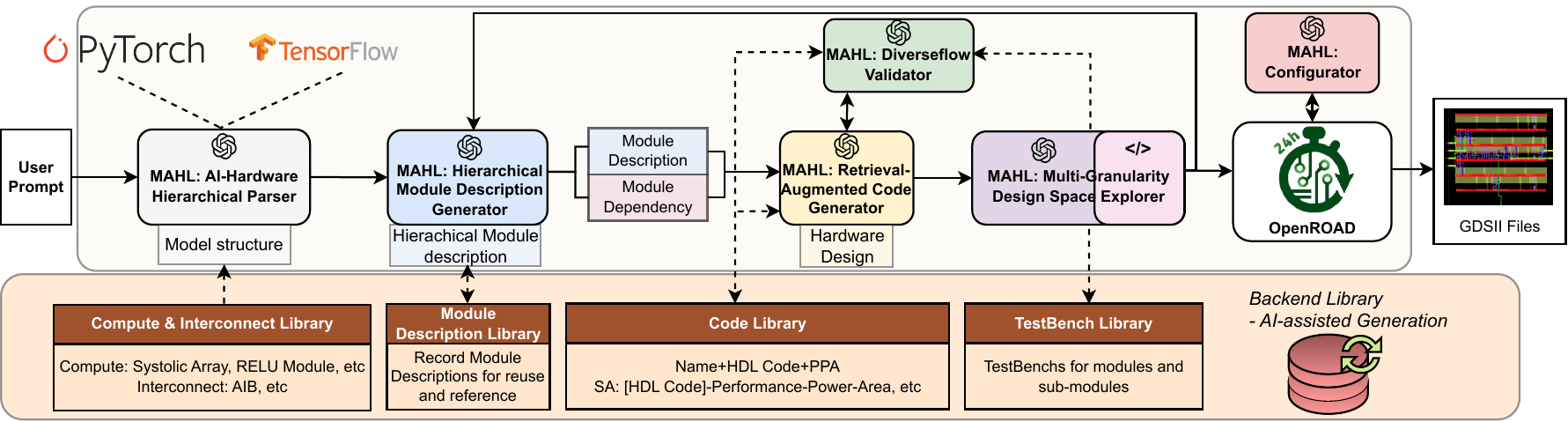} 
    \vspace{-20pt}
    \caption{The overview of the MAHL framework. MAHL framework consists of 6 agents, including an AI-Hardware Hierarchical Parser, a Hierarchical Description Generator, a Retrieval-Augmented Code Generator, an Adaptive Validator, a Multi-granularity Design Space Explorer and a Configurator.}
    \label{fig2}
    \vspace{-20pt}
\end{figure*}

\section{Framework}
\label{sec:fram}

\subsection{Overview}

Figure \ref{fig2} depicts a comprehensive overview of the workflow in our~\name~framework. Leveraging natural language user input specifying the AI algorithm,~\name~automatically extracts neural network layer information, searches and designs the best-fit hardware design structure, and finally implements a chiplet IP with optimized configuration and PPA, aiming to achieve the chiplet design tailored for the algorithm while satisfying user-defined objectives. 
The framework incorporates six agents, each playing a critical role in automating the hardware IP generation and reducing manual design burden in the typical chiplet design workflow: 

(1) The \textbf{AI-Hardware Hierarchical Parser} decomposes user-defined algorithm into multiple computing and interconnection modules, and selects the most appropriate hardware module implementation descriptions by leveraging LLMs and a Compute \& Interconnect library. For components absent from the library, it further incorporates a human-computer interaction (HCI) interface to facilitate real-time completion, enabling a seamless software-to-hardware mapping process;

(2) The \textbf{Hierarchical Module Description Generator} primarily retrieves hierarchical descriptions from the Module Description Library. In cases of retrieval failure, it leverages the provided module placeholders to reconstruct the hierarchy from flattened descriptions. The duo-agent mechanism is also applied to ensure format compliance and content refinement of the generated hierarchical description;

(3) The \textbf{Retrieval-Augmented Code Generator} splits the hierarchical module description into multiple modules and generates corresponding HDL codes in a hierarchical order. To improve code reuse, a dynamically updated Code Library is maintained for high-quality code retrieval;

(4) The \textbf{Diverseflow Validator} integrates conventional simulation and synthesis tools with a multi-round debugging strategy. It generates testbenches using LLMs and retrieves existing ones from the Testbench Library. To enhance output diversity and avoid stagnation on individual failure cases, it introduces controlled noise into prompts, thereby improving validation efficiency;
 
(5) The \textbf{Multi-Granularity Design Space Explorer} receives the submodule PPAs and performs a multi-granularity DSE. It includes both coarse-grained LLM-driven DSE to expand design space and fine-grained analytical DSE for refinement, with the goal of obtaining optimized configurations. After that, the parameters will be fed back to Module Description Generator for full design generation;

(6) The Configurator leverages LLMs to automatically explore and determine the optimized layout-level configuration for physical design generated with OpenROAD \cite{openroad}.

 \begin{figure*}[t]
     \centering
\includegraphics[width=1.0\linewidth]{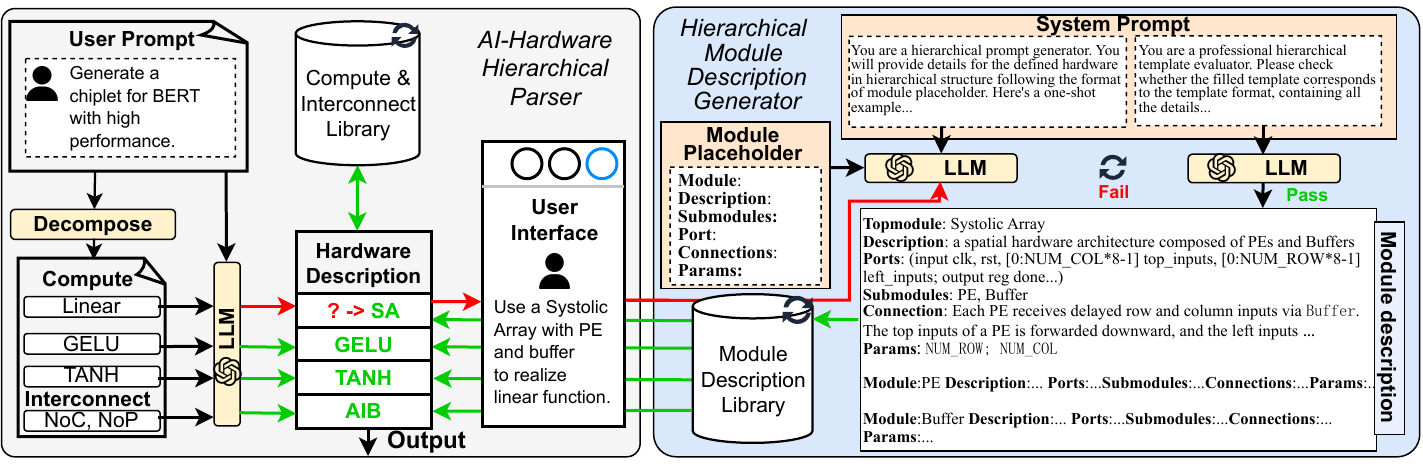} 
    \vspace{-20pt}
    \caption{Hierarchical description generation flow incorporates (1) AI-Hardware Hierarchical Parser and (2) Hierarchical Module Description Generator.}
    \label{fig3}
    \vspace{-15pt}
\end{figure*}

\subsection{Hierarchical Description Generation}

As discussed in Section~\ref{sec:rw}, \textbf{Phase I} of the typical flow focuses on defining the design specification for the target algorithm. Leveraging the modular nature of chiplet-based architecture, this phase is hierarchically realized by two key agents: (1) the \textbf{AI-Hardware Hierarchical Parser} and (2) the \textbf{Hierarchical Module Description Generator} in Figure \ref{fig2}.

\underline{\textbf{AI-Hardware Hierarchical Parser:}} As illustrated in the left part of Figure~\ref{fig3}, the AI-Hardware Hierarchical Parser starts with a user prompt specifying the target algorithm and corresponding design requirements. This prompt may optionally include user-defined hardware preferences. Otherwise, the agent will automatically infer the most suitable hardware configuration based on the algorithm characteristics.

The parser first decomposes the algorithm into structured computational and interconnection layers sourced from Torchvision~\cite{paszke2019pytorch} or Huggingface~\cite{huggingface_platform} using \textit{print(model)} command, which are then passed to LLMs. Leveraging its natural language understanding and domain knowledge, the LLMs evaluates the functional requirements of each layer and determines the most appropriate hardware modules by referencing a built-in library of computing and interconnection unit name with existing module description. The LLMs then produce a preliminary hardware module description after mapping. For any unmapped or ambiguous components that fails to find the corresponding hardware modules, the agent queries the user through an interactive interface to provide additional specifications. The mapping strategy can be abstracted as:

\begin{equation}
\mathcal{M} = \left\{ \left(l_i,\, h_i \right) \,\middle|\, 
\begin{cases}
h_i = \texttt{LLM}(l_i,\, \mathcal{H}_{\text{lib}}), & \text{if matches} \\
h_i = h_i^{\text{user}}, & \text{otherwise}
\end{cases}
\right\}_{l_i \in \mathcal{L}}
\end{equation}

\noindent where $l_i$ denotes the extracted computing or interconnection layers, and $\mathcal{H}_{\text{lib}}$ represents the compute and interconnect library. $\texttt{LLM}(\cdot)$ refers to a mapping function that infers hardware module $h_i$ for each layer $l_i$ through LLMs, and $h_i^{\text{user}}$ refers to the user-provided specification. The resulting complete set of hardware modules serves as the foundation for hierarchical module description generation in subsequent stages.

\underline{\textbf{Hierarchical Module Description Generator:}} For layers that are successfully mapped, the corresponding hardware modules directly retrieve hierarchical descriptions from a Module Description Library, as shown in the right part of Figure~\ref{fig3}. 
For unmapped layer module $h_i^{\text{user}}$, a duo-agent system composed of two LLMs is employed to generate the corresponding hierarchical description. The first LLM, serving as a generator, fills in a module placeholder template with system prompt containing hierarchy hint, which then produces a structured description containing components including \textit{Module}, \textit{Description}, \textit{Submodules}, \textit{Port}, \textit{Connections} and \textit{Params}. The second LLM, serving as an evaluator, evaluates the generated hierarchical description for format correctness and semantic completeness. If the description is valid, a “template pass” token is issued and the description is stored in the Module Description Library. Otherwise, revision suggestions are provided for iterative refinement.

It is worth noting that this agent is invoked in both \textbf{Phase I} for initial specification and \textbf{Phase III} for final design optimization. In the latter case, the system prompt additionally encodes user-specified PPA targets, and the evaluator assesses whether the generated design can reach these goals through the additional optimization. If not, feedback of other potential techniques, such as clock gating for power or pipelining for performance, is returned to guide further refinement. The hierarchical module descriptions of all layers form a set, which constitutes the input to the subsequent design phase.

\subsection{RTL Implementation and Validation}
\label{sec:retrieval}

\begin{figure}[t!]
     \centering
\includegraphics[width=1.0\linewidth]{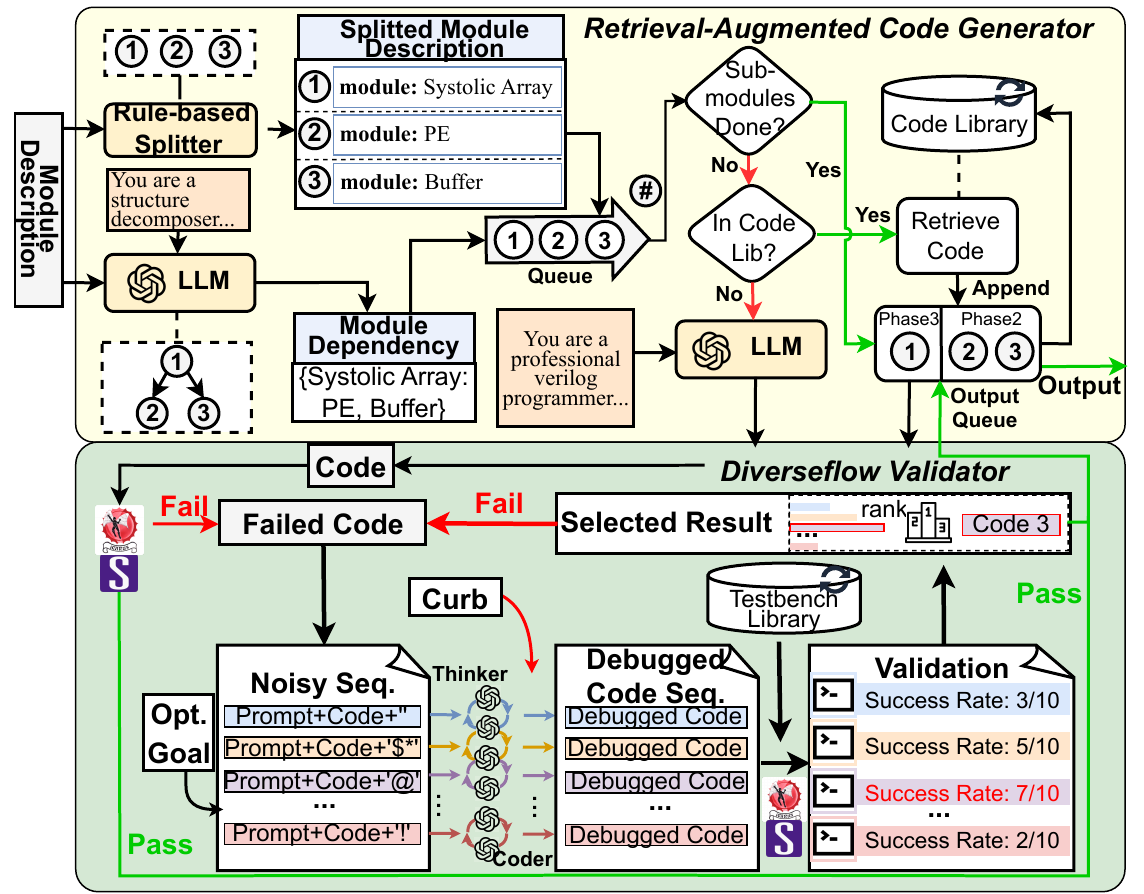} 
    \caption{RTL implementation and validation flow, including (1) Retrieval-Augmented Code Generator and (2) Diverseflow Validator.}
    \label{fig4}
    \vspace{-15pt}
\end{figure}

RTL implementation and validation are involved in both \textbf{Phase II} and \textbf{Phase III}, each focusing on different objectives, as discussed in Section~\ref{sec:rw}. As shown in Figure~\ref{fig2}, this process is supported by two agents: (1) the \textbf{Retrieval-Augmented Code Generator}, and (2) the \textbf{Diverseflow Validator}.

\underline{\textbf{Retrieval-Augmented Code Generator:}} The upper part of Figure~\ref{fig4} illustrates the workflow of Retrieval-Augmented Code Generator, which operates on inputs of hierarchical module descriptions produced by the preceding agents. For each hierarchical module description, the system performs two key steps: a rule-based structural decomposition that segments the design into individual modules, and an LLM-assisted dependency analysis to identify inter-module relationships. The resulting individual module descriptions are then ordered bottom-up according to their dependency hierarchy and placed into a queue for sequential processing.

For designs that have not yet undergone \textbf{Phase III}, the agent first checks whether all submodules specified in the dependency graph have been generated. A complete set of submodules indicates that the RTL code required for behavioral modeling is available, allowing proceeding to the DSE phase. If some submodule codes are still missing, each module description is processed by first querying the dynamically updated Code Library, a repository of HDL code snippets where each entry includes:
\begin{itemize}
    \item \textit{Key}: A unique identifier (typically the module's name or function).
    \item \textit{Weight} $w$: A numerical score reflecting the overall quality and reliability of the code.
    \item \textit{PPA}: Power, performance and area of the code.
    \item \textit{Code}: The corresponding HDL implementation.
\end{itemize}
For each module, the agent retrieves the most relevant entry by computing the cosine similarity between the query token $T$ (derived from the description) and each stored key $T_{db}$:

\begin{equation}
\footnotesize
S_{\max}(T) =\max_{T_{db}\in \mathrm{DB}} ( \cos\left(T, T_{db}\right)).
\end{equation}

while the decision is performed in a two-step check ensuring both the similarity and quality of the retrieved code:
\begin{enumerate}
    \item \textbf{Similarity Check:} If $S_{\max} < t_{\text{sim}}$, where $t_{\text{sim}}$ is a predefined similarity threshold, the retriever directly proceeds to the red fail path to further code generation, as shown in Figure \ref{fig4};
    \item \textbf{Weight Check:} If $S_{\max} \geq t_{\text{sim}}$, the candidate quality is further examined. Only if $w(T_{db}) \geq t_w$, where $t_w$ is the weight threshold, the code will be retrieved and reused (green success path in Figure~\ref{fig4}); otherwise, the retriever opts to the red fail path.
\end{enumerate}
 
When the retriever opts to the red path for new code generation, it invokes the LLM. For any newly generated code, an initial weight $w_0 = \alpha$ is assigned. After undergoing the Diverseflow Validator, the new entry including extracted \textit{Key}, initial weight $w_0$, \textit{Power, Performance (Clock Frequency) and Area (PPA)} values derived from synthesis tools and the validated \textit{Code}, is appended into the Code Library.

For designs that have entered \textbf{Phase III} and are undergoing final RTL code generation, the agent will further dynamically adjust the weights of code entries based on final code validation outcomes. Instead of an explicit equation, this dynamic update is performed via a weight management algorithm (detailed in {Algorithm 1}). In essence, for each code snippet that passes its tests, its weight is multiplied by a factor \(\beta > 1\); if it fails, its weight is reduced by multiplying by \(1/\beta\). Code entries with weights falling below a threshold \(t_h\) are removed from the Code Library through a garbage collection process.

\begin{algorithm}[t]
\footnotesize	
\caption{Dynamic Weight Management for Code Library}
\label{alg:dynamic_weight_update}
\begin{algorithmic}[1]

\REQUIRE Each code block's initial weight \(w_n\) is initialized.
\STATE \(N\): the number of submodules in simulated code
\STATE Set scaling factor \(\beta > 1\)
\FOR{\(n = 1\) to \(N\)}
    \IF{code block \(n\) passes simulation test}
        \STATE \(w_n \gets \beta \cdot w_n\)
    \ELSE
        \STATE \(w_n \gets \frac{1}{\beta} \cdot w_n\)
    \ENDIF
\ENDFOR
\STATE Remove all code blocks with \(w_n < t_h\) (garbage collection)
\end{algorithmic}
\end{algorithm}


\underline{\textbf{Diverseflow Validator:}} The lower part of Figure~\ref{fig4} presents the validation workflow for newly generated code. Each code snippet is first verified for functional correctness using a simulator (e.g., ICARUS Verilog) and a Testbench Library containing validated testbenches. To ensure its reliability, the testbenches are generated manually with assistance from LLMs. Since each module's behavior and interfaces are explicitly defined by the hierarchical prompt, testbench generation can be performed in parallel with code generation, with system prompt such as: ``You are a testbench generator. Please generate testbench for {module description}." Upon passing testbench validation through simulation, the code is synthesized using a standard synthesis tool (e.g., Design Compiler) to extract its PPA metrics and appended to the output queue (green path in Figure \ref{fig4}). The verified code and its corresponding PPA are then stored in the Code Library for future reuse.

For failed code, the agent creates a noisy prompt sequence containing \textit{K} parallel threads. A subset of symbolically represented tokens (comprising \textit{P\%} of the total code length) is selected as noise. The introduction of noise during generation aims to reduce the likelihood of producing similar outputs repeatedly, thereby increasing the chances of generating correct code. Symbolic tokens are specifically chosen as noise to avoid the accidental formation of meaningful words, which could undesirably bias the model's generation. Each of the \textit{K} threads is processed by a duo-agent system, consisting of:
    \begin{itemize}
        \item \textbf{Thinker:} Evaluates the buggy code, diagnosing issues and suggesting specific fixes.
        \item \textbf{Coder:} Applies the suggested modifications without access to the original prompt, ensuring unbiased corrections.
    \end{itemize}

The newly generated debugged code from all \textit{K} threads is then re-executed for simulation and synthesis. The agent sorts and ranks these variants to identify the best-performing code, which refers to the functionally correct RTL implementation that best fits the user-defined optimization objectives in Algorithm \ref{alg:code_selection}, using their designed testbench. At the same time, the validator will record the errors reported from the simulator. If none of the variants fully pass the simulation, the agent will begin another iteration with the failed code with the fewest errors. 

To avoid infinite loops, a \emph{debugging curb} monitors the iteration count \textit{C}. If the number of iterations exceeds \textit{C}, the system halts and requests a human-written debugging manual for subsequent attempts. The human-written debugging manual should point out what's wrong in the code and that response will be attached to the prompt for the debugging code generation from this point on. Upon successful debugging, the best-performing code follows the green path in Figure \ref{fig4} and is placed into the output queue. This RTL implementation and validation process iterates until all required submodules have been retrieved or generated in \textbf{Phase II}, or until all chiplet components with full RTL implementation have been completed in \textbf{Phase III}.

\begin{algorithm}[t]
\footnotesize	
\caption{Code Selection Strategy}
\label{alg:code_selection}
\begin{algorithmic}[1]
\STATE \textbf{Input:} $debugged\_list$: list of debugged code sequences
\STATE \textbf{Input:} $target\_result$: user-defined optimization metric (e.g., $1/\text{clk\_freq}$, power, or area)
\STATE \textbf{Initialize:} $pass\_list \gets [\,]$, $fail\_list \gets [\,]$

\FOR{$i$ in $debugged\_list$}
    \IF{$\text{codes}[i]$ passes simulation}
        \STATE Append $(i,~\text{target\_result}(\text{codes}[i]))$ to $pass\_list$
    \ELSE
        \STATE Append $(i,~\text{num\_failed\_cases}(\text{codes}[i]))$ to $fail\_list$
    \ENDIF
\ENDFOR

\IF{$pass\_list$ is not empty}
    \STATE index $\gets$ index in $pass\_list$ with minimum target result
\ELSE
    \STATE index $\gets$ index in $fail\_list$ with minimum failed cases
\ENDIF

\RETURN codes[index]
\end{algorithmic}

\end{algorithm}

\subsection{Design Space Exploration}
Design Space Exploration is one of the major steps in \textbf{Phase III}, as illustrated in Section \ref{sec:rw}. After collecting the necessary submodule PPAs, the flow will proceed to another agent, called Multi-Granularity Design Space Explorer (Figure \ref{fig2}), to derive the optimal hardware parameters for the input algorithm.

\underline{\textbf{Multi-Granularity Design Space Explorer:}}
In a typical chiplet design flow, DSE poses a significant challenge to the design flow, which is mainly attributed to the extensive design space. Given $i$ parameters, even if each parameter is limited to only 10 possible values, the total number of configurations reaches $10^i$, leading to long runtime. Therefore, designers often use heuristic baseline configurations to find suboptimal but feasible solutions. On the other hand, relying solely on LLMs can compromise precision. To address this challenge, this paper proposes a Multi-Granularity Design Space Explorer, where we integrate the strengths of both LLM-based DSE and analytical DSE.

\begin{figure}[t]
     \centering
\includegraphics[width=1.0\columnwidth]{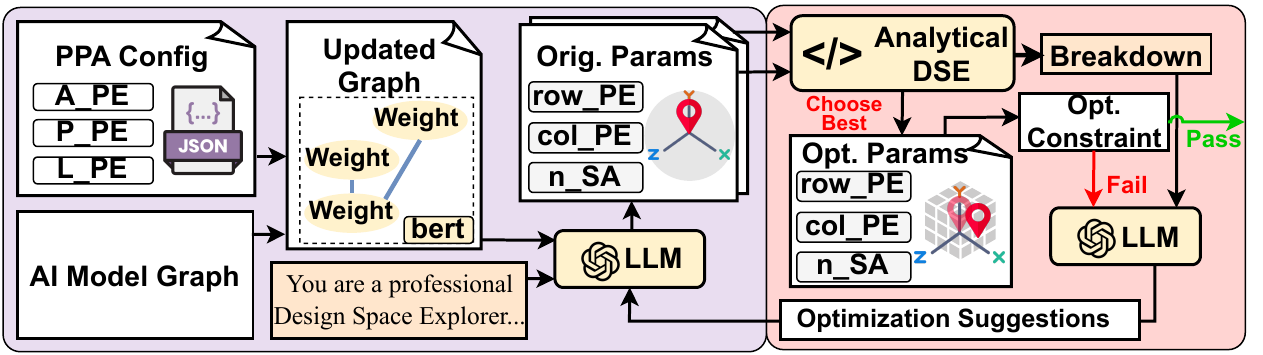} 
    \vspace{-18pt}
    \caption{Workflow of Multi-Granularity Design Space Explorer. }
    \label{fig6}
    \vspace{-20pt}
\end{figure}

Provided in Figure \ref{fig6}, the PPA metrics of submodules are first used to update the node and edge weight in the extracted AI Model Graph, which represents the estimated computing and interconnection overhead of each layer. Given the AI-to-hardware simulation information, LLMs then function as a design space explorer and leverage its domain knowledge to derive $M$ sets of coarse-grained configuration baseline from the whole design space. A precise analytical DSE then conducts refinement by iteratively adjusting the parameters within a limited range based on the coarse-grained configurations, which finally identifies one set of optimal configuration from $M$ sets under the constrained design space.

After selecting the best configuration set, we first evaluate whether it satisfies the predefined constraints, including all PPA (Power, Performance, Area) targets. If the constraints are met, the configuration is accepted and output via the green path. Otherwise, we perform a breakdown of the analytical metrics into computing and interconnection components to identify the performance bottleneck. This bottleneck, along with its associated parameters, is then provided to the LLMs, which generates optimization suggestions for the next DSE iteration. Upon completion of the DSE process, the flow proceeds to the full chiplet RTL implementation, followed by \textbf{Phase IV}, which involves the final layout design.
\underline{\textbf{Layout Configurator:}}
In \textbf{Phase IV}, the final layout is implemented using OpenROAD \cite{openroad}. To generate GDSII files, an LLM-based agent, \textbf{Configurator}, automatically produces and iteratively revises configuration files by parsing the OpenROAD tutorial and error messages. It tunes parameters such as wire length, spacing, chip size, and congestion tolerance, ensuring successful execution of the layout flow.

\section{Evaluation}

\begin{table*}[ht]
    \centering
    \caption{Generation accuracy comparison on LLM-generated hierarchical prompt over 6 designs.}
    \vspace{-10pt}
    \renewcommand{\arraystretch}{1.1}
    \setlength{\tabcolsep}{4pt}
    \resizebox{\textwidth}{!}{%
    \begin{tabular}{cc|cccccc|cccccc|cccccc}
        \hline
        && \multicolumn{6}{c|}{\textbf{GPT-4o}} 
        & \multicolumn{6}{c|}{\textbf{Llama3.3:70b}} 
        & \multicolumn{6}{c}{\textbf{Gemma3.3:27b}} \\
        \rowcolor{lightgray}
        && \textbf{N} & \textbf{H} & \textbf{HR} & \textbf{ND} & \textbf{HD} & \textbf{HRD}
        & \textbf{N} & \textbf{H} & \textbf{HR} & \textbf{ND} & \textbf{HD} & \textbf{HRD}
        & \textbf{N} & \textbf{H} & \textbf{HR} & \textbf{ND} & \textbf{HD} & \textbf{HRD} \\
        \hline
        \textbf{Mux 64to1}         & \textbf{Pass@1} & \cellcolor[HTML]{C3E6CB}1.00 & \cellcolor[HTML]{C3E6CB}1.00 & \cellcolor[HTML]{C3E6CB}1.00 & \cellcolor[HTML]{C3E6CB}1.00 & \cellcolor[HTML]{C3E6CB}1.00 & \cellcolor[HTML]{C3E6CB}1.00 & \cellcolor{gray!20}0.90 & \cellcolor[HTML]{C3E6CB}1.00 & \cellcolor[HTML]{C3E6CB}1.00 & \cellcolor[HTML]{C3E6CB}1.00 & \cellcolor[HTML]{C3E6CB}1.00 & \cellcolor[HTML]{C3E6CB}1.00 &\cellcolor{gray!20} 0.40 & \cellcolor[HTML]{C3E6CB}1.00 & \cellcolor[HTML]{C3E6CB}1.00 & 0.80 & \cellcolor[HTML]{C3E6CB}1.00 & \cellcolor[HTML]{C3E6CB}1.00 \\
                 & \textbf{Pass@5} & \cellcolor[HTML]{C3E6CB}1.00 & \cellcolor[HTML]{C3E6CB}1.00 & \cellcolor[HTML]{C3E6CB}1.00 & \cellcolor[HTML]{C3E6CB}1.00 & \cellcolor[HTML]{C3E6CB}1.00 & \cellcolor[HTML]{C3E6CB}1.00 & \cellcolor{gray!20}0.99 & \cellcolor[HTML]{C3E6CB}1.00 & \cellcolor[HTML]{C3E6CB}1.00 & \cellcolor[HTML]{C3E6CB}1.00 & \cellcolor[HTML]{C3E6CB}1.00 & \cellcolor[HTML]{C3E6CB}1.00 &\cellcolor{gray!20} 0.92 & \cellcolor[HTML]{C3E6CB}1.00 & \cellcolor[HTML]{C3E6CB}1.00 & 0.99 & \cellcolor[HTML]{C3E6CB}1.00 & \cellcolor[HTML]{C3E6CB}1.00 \\
        \hline
        \textbf{Adder 64bit}         & \textbf{Pass@1} & \cellcolor[HTML]{C3E6CB}1.00 & \cellcolor[HTML]{C3E6CB}1.00 & \cellcolor[HTML]{C3E6CB}1.00 & \cellcolor[HTML]{C3E6CB}1.00 & \cellcolor[HTML]{C3E6CB}1.00 & \cellcolor[HTML]{C3E6CB}1.00 &\cellcolor{gray!20} 0.60 & \cellcolor[HTML]{C3E6CB}1.00 & \cellcolor[HTML]{C3E6CB}1.00 & 0.80 & \cellcolor[HTML]{C3E6CB}1.00 & \cellcolor[HTML]{C3E6CB}1.00 &\cellcolor{gray!20} 0.40 & \cellcolor[HTML]{C3E6CB}1.00 & \cellcolor[HTML]{C3E6CB}1.00 & \cellcolor[HTML]{C3E6CB}1.00 & \cellcolor[HTML]{C3E6CB}1.00 & \cellcolor[HTML]{C3E6CB}1.00 \\
                & \textbf{Pass@5} & \cellcolor[HTML]{C3E6CB}1.00 & \cellcolor[HTML]{C3E6CB}1.00 & \cellcolor[HTML]{C3E6CB}1.00 & \cellcolor[HTML]{C3E6CB}1.00 & \cellcolor[HTML]{C3E6CB}1.00 & \cellcolor[HTML]{C3E6CB}1.00 &\cellcolor{gray!20} 0.98 & \cellcolor[HTML]{C3E6CB}1.00 & \cellcolor[HTML]{C3E6CB}1.00 & 0.99 & \cellcolor[HTML]{C3E6CB}1.00 & \cellcolor[HTML]{C3E6CB}1.00 &\cellcolor{gray!20} 0.92 & \cellcolor[HTML]{C3E6CB}1.00 & \cellcolor[HTML]{C3E6CB}1.00 & \cellcolor[HTML]{C3E6CB}1.00 & \cellcolor[HTML]{C3E6CB}1.00 & \cellcolor[HTML]{C3E6CB}1.00 \\
        \hline
        \textbf{Decoder 5to32}         & \textbf{Pass@1} &\cellcolor{gray!20} 0.60 & \cellcolor[HTML]{C3E6CB}1.00 & \cellcolor[HTML]{C3E6CB}1.00 & \cellcolor[HTML]{C3E6CB}1.00 & \cellcolor[HTML]{C3E6CB}1.00 & \cellcolor[HTML]{C3E6CB}1.00 &\cellcolor{gray!20} 0.80 & 0.90 & \cellcolor[HTML]{C3E6CB}1.00 & 0.90 & \cellcolor[HTML]{C3E6CB}1.00 & \cellcolor[HTML]{C3E6CB}1.00 & \cellcolor{gray!20}0.00 & 0.50 &\cellcolor[HTML]{C3E6CB} 0.60 & \cellcolor{gray!20}0.00 & 0.50 &\cellcolor[HTML]{C3E6CB} 0.60 \\
                 & \textbf{Pass@5} &\cellcolor{gray!20} 0.98 & \cellcolor[HTML]{C3E6CB}1.00 & \cellcolor[HTML]{C3E6CB}1.00 & \cellcolor[HTML]{C3E6CB}1.00 & \cellcolor[HTML]{C3E6CB}1.00 & \cellcolor[HTML]{C3E6CB}1.00 &\cellcolor{gray!20} 0.99 & 0.99 & \cellcolor[HTML]{C3E6CB}1.00 & 0.99 & \cellcolor[HTML]{C3E6CB}1.00 & \cellcolor[HTML]{C3E6CB}1.00 & \cellcolor{gray!20}0.00 & 0.96 &\cellcolor[HTML]{C3E6CB} 0.98 & \cellcolor{gray!20}0.00 & 0.96 &\cellcolor[HTML]{C3E6CB} 0.98 \\
        \hline
        \textbf{Barrel Shifter 32bit}         & \textbf{Pass@1} &\cellcolor{gray!20} 0.80 & \cellcolor[HTML]{C3E6CB}1.00 & \cellcolor[HTML]{C3E6CB}1.00 & 0.90 & \cellcolor[HTML]{C3E6CB}1.00 & \cellcolor[HTML]{C3E6CB}1.00 & 0.20 & \cellcolor{gray!20}0.00 & \cellcolor[HTML]{C3E6CB}0.40 & 0.30 & \cellcolor{gray!20}0.00 & \cellcolor[HTML]{C3E6CB}0.40 & \cellcolor{gray!20}0.00 & \cellcolor{gray!20}0.00 & 0.10 & \cellcolor{gray!20}0.00 & \cellcolor{gray!20}0.00 &\cellcolor[HTML]{C3E6CB} 0.50 \\
                & \textbf{Pass@5} &\cellcolor{gray!20} 0.99 & \cellcolor[HTML]{C3E6CB}1.00 & \cellcolor[HTML]{C3E6CB}1.00 & 0.99 & \cellcolor[HTML]{C3E6CB}1.00 & \cellcolor[HTML]{C3E6CB}1.00 & 0.67 & \cellcolor{gray!20}0.00 & \cellcolor[HTML]{C3E6CB}0.92 & 0.83 & \cellcolor{gray!20}0.00 & \cellcolor[HTML]{C3E6CB}0.92 & \cellcolor{gray!20}0.00 & \cellcolor{gray!20}0.00 & 0.41 & \cellcolor{gray!20}0.00 & \cellcolor{gray!20}0.00 &\cellcolor[HTML]{C3E6CB} 0.97 \\
        \hline
        \textbf{Systolic Array 4x4}         & \textbf{Pass@1} & \cellcolor{gray!20}0.00 & 0.30 & 0.60 & 0.30 & 0.50 & \cellcolor[HTML]{C3E6CB}0.70 & \cellcolor{gray!20}0.40 & 0.50 &\cellcolor[HTML]{C3E6CB} 0.80 & 0.70 & 0.60 &\cellcolor[HTML]{C3E6CB} 0.80 & \cellcolor{gray!20}0.00 & \cellcolor{gray!20}0.00 &\cellcolor[HTML]{C3E6CB} 0.20 & \cellcolor{gray!20}0.00 & \cellcolor{gray!20}0.00 & \cellcolor[HTML]{C3E6CB}0.20 \\
                 & \textbf{Pass@5} & \cellcolor{gray!20}0.00 & 0.83 & 0.98 & 0.83 & 0.96 & \cellcolor[HTML]{C3E6CB}0.99 & \cellcolor{gray!20}0.92 & 0.96 &\cellcolor[HTML]{C3E6CB} 0.99 & 0.97 & 0.98 &\cellcolor[HTML]{C3E6CB} 0.99 & \cellcolor{gray!20}0.00 & \cellcolor{gray!20}0.00 &\cellcolor[HTML]{C3E6CB} 0.67 & \cellcolor{gray!20}0.00 & \cellcolor{gray!20}0.00 & \cellcolor[HTML]{C3E6CB}0.67 \\
        \hline
        \textbf{UART 8bit}         & \textbf{Pass@1} & \cellcolor{gray!20}0.00 & \cellcolor[HTML]{C3E6CB}1.00 & \cellcolor[HTML]{C3E6CB}1.00 & 0.20 & \cellcolor[HTML]{C3E6CB}1.00 & \cellcolor[HTML]{C3E6CB}1.00 & \cellcolor{gray!20}0.60 & 0.70 & \cellcolor[HTML]{C3E6CB}1.00 & 0.70 & \cellcolor[HTML]{C3E6CB}1.00 & \cellcolor[HTML]{C3E6CB}1.00 & 0.50 &\cellcolor{gray!20} 0.40 &\cellcolor[HTML]{C3E6CB} 0.80 & 0.60 & 0.50 &\cellcolor[HTML]{C3E6CB} 0.80 \\
                 & \textbf{Pass@5} & \cellcolor{gray!20}0.00 & \cellcolor[HTML]{C3E6CB}1.00 & \cellcolor[HTML]{C3E6CB}1.00 & 0.67 & \cellcolor[HTML]{C3E6CB}1.00 & \cellcolor[HTML]{C3E6CB}1.00 & \cellcolor{gray!20}0.98 & 0.99 & \cellcolor[HTML]{C3E6CB}1.00 & 0.99 & \cellcolor[HTML]{C3E6CB}1.00 & \cellcolor[HTML]{C3E6CB}1.00 & 0.97 &\cellcolor{gray!20} 0.92 &\cellcolor[HTML]{C3E6CB} 0.99 & 0.98 & 0.97 &\cellcolor[HTML]{C3E6CB} 0.99 \\
        \hline
    \end{tabular}
    }
\label{table1}
\vspace{-13pt}
\end{table*}

\subsection{Experimental Setup}
\textbf{Dataset.}
Dataset I is composed of simple designs, including the Multiplexer, the Adder, the Decoder, the Barrel Shifter, the Systolic Array, the AES Block Cipher and the UART. These designs are the same as the first dataset in ROME~\cite{ROME} and are used to test the generation and validation ability of the Retrieval-Augmented Code Generator and Diverseflow Validator. Given the focus of this study on chiplets, in Dataset II, we select and evaluate chiplet designs for different AI algorithms. Typically, their components involve a Systolic Array with PEs and buffers to implement 'Conv1D'/'Linear'/'Conv2D' layers and activation modules to implement other activation layers. For interconnection, we employ channels of the AIB 2.0 interface \cite{AIB} for the NoP while adopting multiple links per channel with 8 bits for NoC interface to ensure same bandwidth with NoP. We use AI model, including BERT, LLaMA and GPT for algorithm-hardware mapping.

\textbf{Configuration and Platforms.} Our~{\name} employs three LLM models to perform RTL hierarchy-aware design generation and validation. We use one closed-source model, GPT-4o\cite{gpt4}, and two open-source models, LLaMA 3.3-70B\cite{meta_llama3} and Gemma 3-27B\cite{gemma3}. For the GPT-4o model, we configure the temperature to 0.8 to approximate the behavior of its online version. For the local models, we use a temperature of 0.6, following the default setting in the Ollama library. 

The Diverseflow Validator is configured with $K=2$ threads, one with noise and one without noise. We define the curb count to be $C=5$ and noise percentage to be $P=30\%$. To ensure a fair comparison and save execution time, we only conduct 1 iteration of debugging for simple design, while for chiplet design, we set the debugging iterations to be 5. For the Multi-Granularity Design Space Explorer, we search $M=80$ sets of coarse-grained configurations to enable comparable numbers of design points to CLARIE (expert-based) \cite{CLARIE}. For Section \ref{sec:simple}, we conduct 10 trials to measure Pass Rate, while for Section \ref{sec:chiplet}, the number of trials is set to 20.

\begin{figure}[t]
     \centering
\includegraphics[width=0.9\columnwidth]{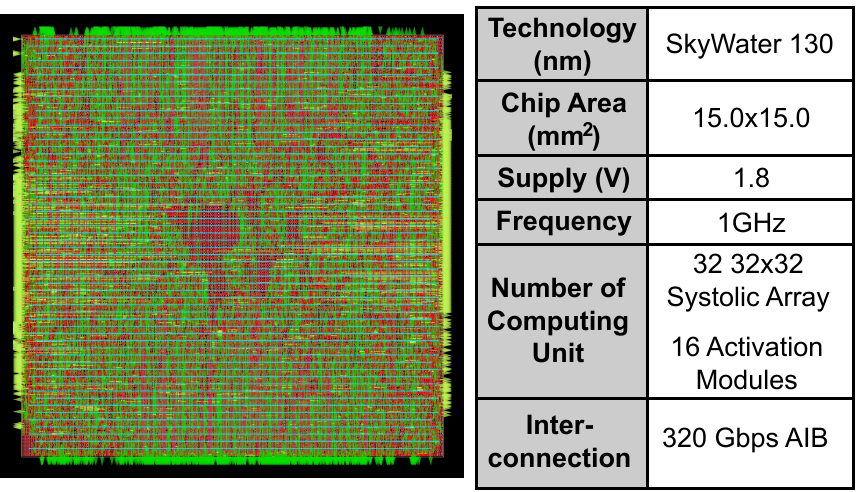} 
    \vspace{-5pt}
    \caption{Layout of one chiplet design for BERT algorithm. }
    \label{fig8}
    \vspace{-15pt}
\end{figure}

For the simulation EDA tools, we integrate the ICARUS Verilog \cite{iverilog} to automate the workflow. For hardware implementations that exist in the Compute \& Interconnect Library, we use the same cycle-accurate chiplet simulator as CLARIE when conducting analytical DSE. The LLM-aided hardware design is synthesized based on a TSMC 28nm technology using Design Compiler to enable fair comparison with CLARIE for PPA comparison in Section \ref{sec:chiplet}. For the layout design, we resynthesize one chiplet and conduct complete layout procedure under SkyWater 130nm with OpenROAD \cite{openroad}, which is for the BERT model under compact area mode. The chiplet is composed of 32 32x32 Systolic Array, 16 Activation Units (GELU and TANH modules), connected through the 320 Gbps AIB channels, as shown in Figure \ref{fig8}. Most experiments are run on a Linux server with 4×A6000 graphics cards and a CPU of AMD EPYC 7763 64-Core Processor.  

\label{sec:eval}

\textbf{Experimental Metric.}  We evaluate the experimental results using multiple metrics. For generation accuracy, we apply Pass@k metrics \cite{ROME} to assess the generation pass rate:
\begin{equation}
\text{pass@k} = \mathbb{E}_{\text{Problems}} \left[ 1 - \frac{\binom{n - c_p}{k}}{\binom{n}{k}} \right]
\end{equation}
where $n$ is the total number of generations, $c_p$ is the number of successes, and $k$ is the number of attempts considered. We apply $Pass@1$ and $Pass@5$ in our experiment. We also evaluate the generation quality, i.e., PPA optimization, based on metrics including power, latency and area.
\begin{table}[t!]
    \centering
    \caption{Comparison of optimized DSE result from Multi-Granularity Design Space Explorer with GPT-4o with DSE from baseline CLARIE (\cite{CLARIE}), a human expert experience dependent framework. The bandwidth of both NoC and NoP $BW_{NoC/NoP}$ is measured in Gbps.}
    \renewcommand{\arraystretch}{1.0}
    \setlength{\tabcolsep}{4pt}

    \textbf{(a) High Performance Mode.}
    \vspace{0.2cm}
    \resizebox{\columnwidth}{!}{%
    
    \begin{tabular}{c|ccc|ccc}
     
        \hline
        & \multicolumn{3}{c|}{\textbf{CLARIE (\cite{CLARIE})}} & \multicolumn{3}{c}{\textbf{MAHL}} \\

        &  BERT & LLaMA & GPT  & BERT & LLaMA & GPT    \\
        \hline
        size\textsubscript{SA} & 32$\times$64  & 32$\times$64 & 32$\times$64 & 32$\times$64  & 32$\times$64 & 32$\times$64   \\
        n\textsubscript{SA} &  128  & 128 &  128& 64  & 128 &  128    \\
        type\textsubscript{act} & GELU, TANH  & SILU & GELU  & GELU, TANH  & SILU & GELU   \\
  
        n\textsubscript{act} & 64  & 64 &  64 &  64  & 64 &  64  \\
        BW\textsubscript{NoC/NoP} & 320 & 320 & 320 & 320 & 320 & 320\\
        \hline
    \end{tabular}
    }
     \vspace{0.2cm}
    \textbf{(b) Compact Area Mode.}
    \vspace{0.2cm}
    \resizebox{\columnwidth}{!}{%
    
   \begin{tabular}{c|ccc|ccc}
    \hline
    & \multicolumn{3}{c|}{\textbf{CLARIE (\cite{CLARIE})}} & \multicolumn{3}{c}{\textbf{MAHL}} \\

    &  BERT & LLaMA & GPT  & BERT & LLaMA & GPT   \\
    \hline
    size\textsubscript{SA} & 32$\times$32  & 32$\times$32 & 32$\times$32 & 32$\times$32  & 16$\times$16 & 32$\times$32   \\
    n\textsubscript{SA} &  32  & 32 &  32 & 32  & 16 &  32    \\
    type\textsubscript{act} & GELU, TANH  & SILU& GELU  & GELU, TANH  & SILU & GELU  \\

    n\textsubscript{act} & 16  & 16 &  16 &  16  & 8 &  16  \\
            BW\textsubscript{NoC/NoP} & 320 & 320 & 320 & 320 & 320 & 320\\
    \hline
\end{tabular}
}
\label{table3}
\vspace{-18pt}
\end{table}
\subsection{Simple Design Generation with MAHL}
\label{sec:simple}

As shown in Table \ref{table1}, the simple design generation with MAHL is composed of 6 different metrics as follows:

\begin{itemize}
    \item \textbf{N} stands for conventional LLM generation with \textbf{N}on-hierarchical prompt;
    \item \textbf{H} stands for \textbf{H}ierarchical generation with Hierarchical Module Description Generator alone;
    \item \textbf{HR} stands for \textbf{H}ierarchical Generation with \textbf{R}etrieval-Augmented Code Generator alone;
    \item \textbf{ND} stands for \textbf{N}on-hierarchical generation with \textbf{D}iverseflow Validator alone;
    \item \textbf{HD} stands for \textbf{H}ierarchical Generation with \textbf{D}iverseflow Validator alone;
    \item \textbf{HRD} stands for \textbf{H}ierarchical Generation with \textbf{R}etrieval-Augmented Code Generator and \textbf{D}iverseflow Validator.
\end{itemize}
Note that when we measure the pass rate, we only count it as a success when it passes both the syntax and functional tests. Also, there is no combination of non-hierarchical generation with the Retrieval-Augmented Code Generator, since we do not store the non-hierarchical code for this evaluation.

\textbf{Improvement with Hierarchical Generation:}  
The hierarchical generation approach yields up to an improvement of 0.5 in Pass@1, and provides an average gain of approximately 20\%–40\% in Pass@1 across most cases. Additionally, it enables the integration of the retrieval approach, further enhancing module reuse.

\textbf{Improvement with Diverseflow Validation:}  
The Diverseflow Validator, two parallel threads within a single validation iteration, demonstrates excellent performance. It achieves up to a 0.6 improvement in Pass@1 and also provides consistent gains across other cases.

\textbf{Improvement with Retrieval-Augmented Code Generation:}  
The code retrieval component typically improves the pass rate by 10\%–40\%. It maintains a library of essential sub-modules for each generation, which are pre-verified as described in Section \ref{sec:fram}.

\textbf{Improvement with the Whole MAHL Framework:}  
The overall framework consistently enhances generation performance across all cases, with a maximum improvement of 1.0 in Pass@1. Compared to the Non-Hierarchical Generation baseline, our full design (HRD) achieves an average Pass@1 gain of 44.67\%.

\subsection{Chiplet IP Generation with MAHL}
\label{sec:chiplet}

\begin{table*}[h!]
\centering
\caption{PPA optimization result including Pass@5, energy (uJ), latency (ns), area (mm\textsuperscript{2}) and power density (mW/mm\textsuperscript{2}) compared with general LLMs and three-year hardware design expert (represented by \cite{CLARIE}). `/' represents failure in LLM-based generation or unapplicable for human expert design. Green highlights the best results. We use GPT-4o and Llama-3.3 as the general LLMs.}
\vspace{-5px}
\renewcommand{\arraystretch}{1.0}
\setlength{\tabcolsep}{4pt}
\scriptsize

\textbf{(a) High Performance Mode.}
\vspace{0.1cm}

\begin{tabular}{c|ccccc|ccccc|ccccc}

\hline

& \multicolumn{5}{c|}{\textbf{BERT}} & \multicolumn{5}{c|}{\textbf{LLaMA}} & \multicolumn{5}{c}{\textbf{GPT}} \\
& Pass@5 & Energy & Latency & Area & P.D. & Pass@5 & Energy & Latency & Area & P.D. & Pass@5 & Energy & Latency & Area & P.D. \\
\hline
CLARIE & / & 647.11 & 2812.58 & \cellcolor[HTML]{C3E6CB}182.25 & 1262.43 & / & \cellcolor[HTML]{C3E6CB}34566.05 & 136219.25 & 182.25 & \cellcolor[HTML]{C3E6CB}1392.33 & / & \cellcolor[HTML]{C3E6CB}538.85 & 2220.71 & 182.25 & \cellcolor[HTML]{C3E6CB}1331.40 \\
GPT-4o & 0 & / & / & / & / & 0 & / & / & / & / & 0 & / & / & / & / \\
MAHL (ours) & \cellcolor[HTML]{C3E6CB}0.72 & \cellcolor[HTML]{C3E6CB}470.22 & \cellcolor[HTML]{C3E6CB}1806.54 & 236.81 & \cellcolor[HTML]{C3E6CB}1099.14 
& \cellcolor[HTML]{C3E6CB}0.25 & 49848.65 & \cellcolor[HTML]{C3E6CB}123466.67 & \cellcolor[HTML]{C3E6CB}137.61 & 2933.96 
& \cellcolor[HTML]{C3E6CB}0.60 & 595.78 & \cellcolor[HTML]{C3E6CB}2151.82 & \cellcolor[HTML]{C3E6CB}144.02 & 1922.46 \\
\hline
CLARIE & / & 647.11 & \cellcolor[HTML]{C3E6CB}2812.58 & 182.25 & \cellcolor[HTML]{C3E6CB}1262.43 & / & \cellcolor[HTML]{C3E6CB}34566.04 & \cellcolor[HTML]{C3E6CB}136219.25 & \cellcolor[HTML]{C3E6CB}182.25 & \cellcolor[HTML]{C3E6CB}1392.33 & / & \cellcolor[HTML]{C3E6CB}538.85 & 2220.71 & 182.25 & \cellcolor[HTML]{C3E6CB}1331.40 \\
Llama-3.3:70b & 0 & / & / & / & / & 0 & / & / & / & / & 0 & / & / & / & / \\
MAHL (ours) & \cellcolor[HTML]{C3E6CB}0.60 & \cellcolor[HTML]{C3E6CB}613.81 & 3132.10 & \cellcolor[HTML]{C3E6CB}132.45 & 1479.61 & 0 & / & / & / & / & \cellcolor[HTML]{C3E6CB}0.44 & 702.93 &  \cellcolor[HTML]{C3E6CB}2194.76 &  \cellcolor[HTML]{C3E6CB}172.16 & 1860.34 \\
\hline
\end{tabular}

\vspace{0.2cm}
\textbf{(b) Compact Area Mode.}
\vspace{0.1cm}

\begin{tabular}{c|ccccc|ccccc|ccccc}
\hline
& \multicolumn{5}{c|}{\textbf{BERT}} & \multicolumn{5}{c|}{\textbf{LLaMA}} & \multicolumn{5}{c}{\textbf{GPT}} \\
& Pass@5 & Energy & Latency & Area & P.D. & Pass@5 & Energy & Latency & Area & P.D. & Pass@5 & Energy & Latency & Area & P.D. \\
\hline
CLARIE & / & 647.11 & \cellcolor[HTML]{C3E6CB}12223.84 & 25 & 2117.53 & / & 34566.05 & \cellcolor[HTML]{C3E6CB}682904.83 & 25 & \cellcolor[HTML]{C3E6CB}2024.65 & / & 538.85 & \cellcolor[HTML]{C3E6CB}8956.54 & 25 & \cellcolor[HTML]{C3E6CB}2406.51 \\
GPT-4o & 0 & / & / & / & / & 0 & / & / & / & / & 0 & / & / & / & / \\
MAHL (ours) & \cellcolor[HTML]{C3E6CB}0.60 & \cellcolor[HTML]{C3E6CB}503.78 & 12228.47 & \cellcolor[HTML]{C3E6CB}21.69 & \cellcolor[HTML]{C3E6CB}1899.37 
& 0.25 & 27562.16 & 2660426.90 & \cellcolor[HTML]{C3E6CB}1.61 & 6434.81 
& \cellcolor[HTML]{C3E6CB}0.60 & \cellcolor[HTML]{C3E6CB}399.50 & 11059.74 & \cellcolor[HTML]{C3E6CB}3.62 & 9978.45 \\
\hline
CLARIE & / & 647.11 & \cellcolor[HTML]{C3E6CB}12223.84 & 25 & 2117.53 & / & 34566.04 & \cellcolor[HTML]{C3E6CB}682904.83 & 25 & \cellcolor[HTML]{C3E6CB}2024.65 & / & 538.85 & \cellcolor[HTML]{C3E6CB}8956.54 & 25 & \cellcolor[HTML]{C3E6CB}2406.51 \\
Llama-3.3:70b & 0 & / & / & / & / & 0 & / & / & / & / & 0 & / & / & / & / \\
MAHL (ours) & \cellcolor[HTML]{C3E6CB}0.44 & \cellcolor[HTML]{C3E6CB}501.54 & 41956.70 & \cellcolor[HTML]{C3E6CB}6.23 & \cellcolor[HTML]{C3E6CB}1918.74
& \cellcolor[HTML]{C3E6CB}0.44 & \cellcolor[HTML]{C3E6CB}22869.55 & 816798.56 & \cellcolor[HTML]{C3E6CB}4.26 & 6572.54 
& 0.25 & \cellcolor[HTML]{C3E6CB}453.62 & 9471.36 & \cellcolor[HTML]{C3E6CB}13.14 & 3644.90 \\
\hline
\end{tabular}

\label{table4}
\vspace{-10pt}
\end{table*}


For chiplet IP design, Dataset II specializes its configuration for three AI models, which are BERT, GPT and LLaMA. Besides the same generation and validation function of~{\name} as that used in the simple Verilog design generation, this experiment incorporates the AI-Hardware Hierarchical Parser and the Multi-Granularity Design Space Explorer to realize the whole flow of chiplet design. We apply the same mapping strategy as the one in CLARIE \cite{CLARIE} for fair comparison, which is a chiplet design framework based on simulation and human expert experience. We allow a total of $n=20$ generations and $M=80$ coarse-grained configurations for each design. For the design objective, we choose $high\_performance$ and $compact\_area$ respectively. Other than the user-defined optimization objective mentioned in Section \ref{sec:retrieval}, we also extend the decision making of best results by adding two soft constraints, abstracted as follows:

\begin{equation}
\footnotesize
x^* = \arg\max_{x \in S} O(x), \quad S = X' \text{ if } X' \neq \emptyset \text{ else } X
\end{equation}
\begin{equation}
\footnotesize
    X' = \{ x \in X \mid P_1(x) \geq T_1, P_2(x) \geq T_2 \}
\end{equation}

Here, $P_1(x)$ and $P_2(x)$ represents two soft constraints other than the optimization objective in PPA. $T_1$ and $T_2$ are the thresholds for area and power density. In our experiment, we set $T_1$ to be 250 mm\textsuperscript{2} and $T_2$ to be 2500 mW/mm\textsuperscript{2}.

\begin{figure}[t]
     \centering
\includegraphics[width=0.85\columnwidth]{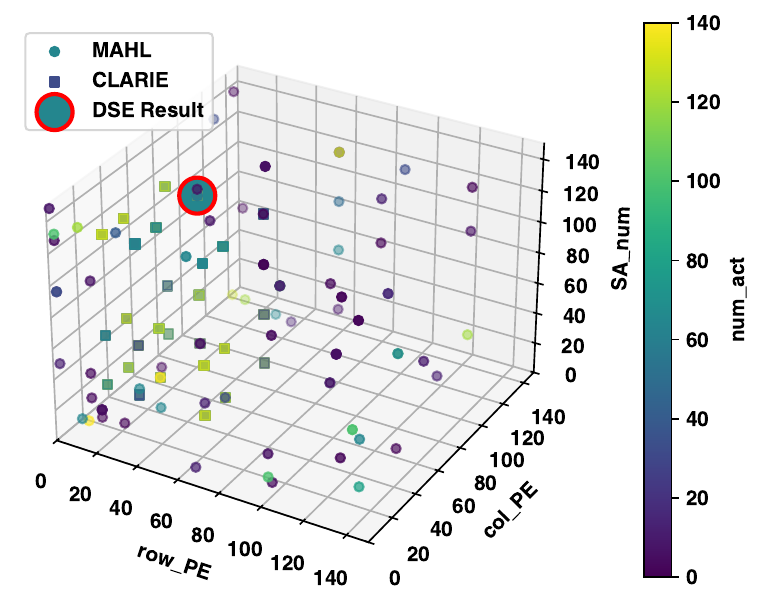} 
    \vspace{-15pt}
    \caption{Design space comparison of the computing parameters between CLARIE (human) and MAHL (ours) on BERT algorithm. }
    \label{fig7}
    \vspace{-17pt}
\end{figure}

As shown in Table \ref{table3}, the configuration is evaluated and found to be close to the golden configurations identified by CLARIE, which is attributed to the integration of LLM-based DSE and analytical DSE without manually identify the baseline. The design space of MAHL is largely extended compared to CLARIE, as shown in Figure \ref{fig7}.

As shown in Table \ref{table4}, our experiments display that utilizing MAHL enhances the generation accuracy compared to conventional LLMs. Specifically, $Pass@5$ rises from 0 to 0.72 for BERT with GPT-4o and 0 to 0.44 for GPT with Llama-3.3 under high performance mode. Even though few of the LLM-generated chiplet designs, such as that for LLaMA under compact area mode, still struggle to reach the level of better trade-off from human experts, the PPA performance of most chiplet designs are already on the same order of magnitude as human expert, demonstrating its great potential for automatic generation and exploration of chiplet design. Insightfully, we observe that LLM-based generation tends to perform better on the primary design objective while underperforming compared to human experts on softer metrics, such as the power density represented by $P_2(x)$. This indicates that the LLM-based framework has yet to achieve the comprehensive trade-off among all design factors that should be taken into account for chiplet design. 

When looking into general LLMs integrated in our MAHL framework, the results reflect that GPT-4o would bring more correct generations and better results on the design objective than Llama-3.3 in most cases, while Llama-3.3 tends to have a better trade-off of all hardware design factors considering $P_1(x)$ and $P_2(x)$. However, compared to using conventional general-purpose LLMs for generation, they both exhibit a significant improvement in success rate, which fails in all cases when generating complex chiplet design.

\section{Conclusion}
\label{sec:conclusion}

In this paper, we propose MAHL, a LLM-driven chiplet generation framework. With integrating promising techniques with LLM generation, MAHL exhibits its potential to be an effective solution to flattened generation, high validation cost and imprecise parameter optimization presented by directly using conventional LLMs. Extensive experiments show that {\name} can reach an improved hardware generation accuracy. Our framework also supports chiplet IP generation with comparable PPA performance with human expert under certain optimization objectives for specific AI models.

\section{Acknowledgment}

\bibliographystyle{IEEEtran}
\bibliography{references,ding}

\end{document}